\documentclass[prl,twocolumn,showpacs,preprintnumbers,amsmath,amssymb,superscriptaddress]{revtex4-1}
\usepackage{amsfonts}
\usepackage{tabularx}
\usepackage{amsmath}
\usepackage{amssymb}
\usepackage{graphicx}
\usepackage{dcolumn}
\usepackage{bm}
\usepackage{ulem}
\usepackage{times}
\usepackage{amsmath, dcolumn, bm, amsthm, amsfonts, amssymb, tabularx, slashbox, subfigure}
\usepackage[colorlinks=true,urlcolor=blue,citecolor=blue,linkcolor=blue]{hyperref}
\usepackage{graphicx,graphics,color}
\def\be{\begin{equation}}
\def\ee{\end{equation}}
\def\bea{\begin{eqnarray}}
\def\eea{\end{eqnarray}}
\def\bse{\begin{subequations}}
\def\ese{\end{subequations}}

\def\be{\begin{eqnarray}}
\def\ee{\end{eqnarray}}

\begin{document}

\title{Topological Fulde-Ferrell and Larkin-Ovchinnikov states in spin-orbit coupled lattice system}
\author{Yao-Wu Guo}
\affiliation{Department of Physics and State Key Laboratory of Surface Physics, Fudan University, Shanghai 200433, China}
\author{Yan Chen}
\affiliation{Department of Physics and State Key Laboratory of Surface Physics, Fudan University, Shanghai 200433, China}
\affiliation{Collaborative Innovation Center of Advanced Microstructures, Nanjing 210093, China}
\date{\today}
\begin{abstract}

The spin--orbit coupled lattice system under Zeeman fields provides an ideal platform to realize exotic pairing states. Notable examples range from the topological superfluid/superconducting (tSC) state, which is gapped in the bulk but metallic at the edge, to the Fulde--Ferrell (FF) state (having a phase-modulated order parameter with a uniform amplitude) and the Larkin--Ovchinnikov (LO) state (having a spatially varying order parameter amplitude). Here, we show that the topological FF state with Chern number ($\mathcal{C}=-1$) (tFF$_{1}$) and topological LO state with $\mathcal{C}=2$ (tLO$_{2}$) can be stabilized in Rashba spin--orbit coupled lattice systems in the presence of both in-plane and out-of-plane Zeeman fields. Besides the inhomogeneous tSC states, in the presence of a weak in-plane Zeeman field, two topological BCS phases may emerge with $\mathcal{C}=-1$ (tBCS$_{1}$) far from half filling and $\mathcal{C}=2$(tBCS$_{2}$) near half filling. We show intriguing effects such as different spatial profiles of order parameters for FF and LO states, the topological evolution among inhomogeneous tSC states, and different non-trivial Chern numbers for the tFF$_{1}$ and tLO$_{1,2}$ states, which are peculiar to the lattice system. Global phase diagrams for various topological phases are presented for both half-filling and doped cases. The edge states as well as local density of states spectra are calculated for tSC states in a 2D strip.

\end{abstract}


\maketitle

% Ming Gong verifed at 7.1.2013, please select several proper codes from the following candidates.
% 03.75.Ss Degenerate Fermi gases
% 74.20.Fg: BCS theory and its development
% 74.70.Tx: Heavy-fermion superconductors
% 74.25.Dw: Superconductivity phase diagrams
% 03.67.Lx Quantum computation architectures and implementations
% 03.65.Vf Phases: geometric; dynamic or topological
% 74.20.-z Theories and models of superconducting state
Topological quantum states of matter have attracted a tremendous amount of attention both theoretically and experimentally during the last decade\cite{Xiao,Qi}. In particular, the topological superfluid/superconducting (tSC) state hosts zero-energy edge states usually related to Majorana fermions (MFs), which are their own antiparticles\cite{Wilczek,majorana}. MFs obey non-Abelian braiding statistics and have potential applications in fault-tolerant topological quantum computation\cite{TQC}; therefore, probing MFs in solid-state and ultracold-atom systems is a topic of considerable interest\cite{Rokhinson,Das,Deng,Mourik,Mao}. At present, a number of exotic quantum systems, such as $\nu =5/2$ fractional quantum Hall states \cite{qhe}, chiral \textit{p}-wave superconductors \cite{srruo,p-wave}, and heterostructures composed of $s$-wave superconductors and semiconductor nanowires or topological insulators \cite{Fu, JSau, Roman,Oreg, Alicea, Lee, Mao, Qi}, are believed to support MFs. Besides these real materials, the experimental realization of spin--orbit coupling (SOC) in cold-atom superfluid systems\cite{soc1,soc2,soc3} provides an ideal platform to search exotic tSC states. In these systems, the SOC and out-of-plane Zeeman field mix different spin states and split the spin states at the Fermi surface, leading to intra-band and inter-band pairings. If the intra-band pairing is an effective \textit{p}-wave interaction, the system becomes topologically non-trivial. Recently, much theoretical work has focused on the tSC state in the presence of the out-of-plane Zeeman field\cite{sato,Fu,Hui2011}. Nonetheless, inhomogeneous superfluid/superconducting states with spontaneously broken translational invariance, known as Fulde--Ferrell (FF) and Larkin--Ovchinnikov (LO) states, are still at the center of interest in many diverse fields of physics \cite{FFLOreview,FFLO2,FFLO3,FFLO1}. Predicted theoretically around 50 years ago \cite{FF64,LO64}, both FF and LO states contain a finite number of Cooper pairs with center-of-mass momentum, arising from the interplay between the competing magnetic and SC orders. In particular, the FF state corresponds to a phase-modulated order parameter of uniform amplitude, whereas the LO state has a spatially varying order parameter amplitude. Normally, the LO state has lower energy than the FF state. Recently, some experimental evidence has been reported in the heavy fermion superconductors \cite{FFLO2} and ultracold Fermi gases \cite{FFLO1}. However, there is still a lack of direct experimental evidence verifying the existence of FF state or LO state.

In the presence of both anisotropic SOC and effective Zeeman fields, recent theoretical studies revealed that the FF state with center-of-mass momentum perpendicular to the direction of the anisotropic SOC can be stabilized in a two-dimensional Fermi gas. A similar route to observe the FF phase in a three-dimensional Fermi gas was proposed\cite{WYi,zheng}. As is well known, by changing the single-particle dispersion, the SOC plays an important role in inducing the topological properties of the superfluid/superconducting system \cite{Qi}. A natural question to ask is whether the FF state may be compatible with topological order. With the recent theoretical and experimental progress on SOC\cite{soc1,soc2,soc3,liu}, $h_{z}$ can make the chemical potential fill the Fermi surface of a single helicity band, and $h_{x}$ deforms the structure of the Fermi surface, so the topologically non-trivial FF state has been predicted in the continuous Fermi gas\cite{cqu,zha,chen,yecao}. In contrast, when the Fermi energy level crosses both helicity bands simultaneously at half filling in the SOC lattice system\cite{zhoutao}, it is possible to contribute to the Cooper pair from two Fermi surfaces; in addition, it is significantly different from the results being obtained by the continuous model. Moreover, this kind of Cooper pair may drive the lattice system into the generalized LO state. For this instance, three questions can be posed. The first is whether the topological phase can persist into the LO phase; the second is, by tuning the chemical potential, can the tFF state emerge in the lattice system? The last is, do they support MFs?

\begin{figure}[h!]
\centering
\resizebox{3.4in}{3.4in}{\includegraphics{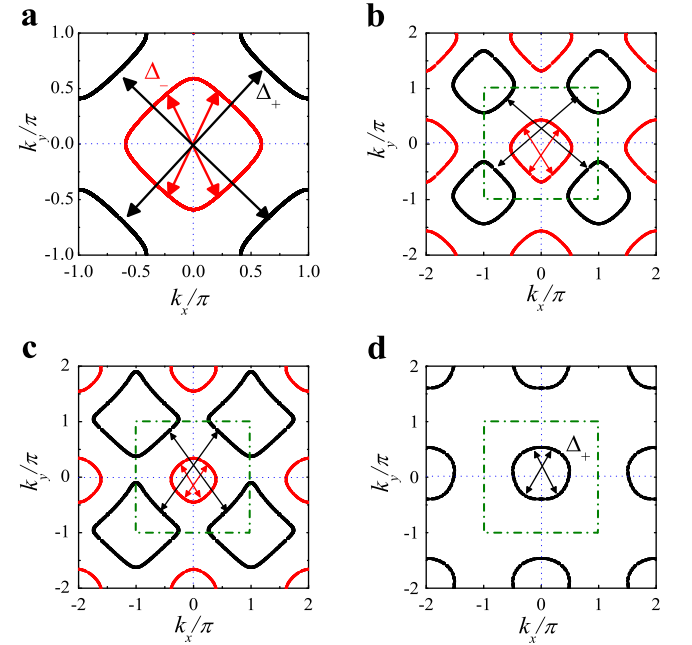}}
\caption{\textbf{Pairing mechanism on the Fermi surface for FF and LO states}: \textbf{(a)} intra-branch BCS pairing in two Fermi surfaces without Zeeman fields, \textbf{(b)} intra-branch pairing (the LO pairing) in two Fermi surfaces with Zeeman fields $h_{z}=h_{x}=0.8$ for half filling, and \textbf{(c)} the doped case with Zeeman fields $h_{z}=1.0$, $h_{x}=0.5$, $\mu=-0.9$. \textbf{(d)} The case of (c), but with $\mu=-4.0$; here, only a single Fermi surface exists.}
\label{fig-FS}
\end{figure}

In this article, we report the existence of the topological non-trivial LO phase with a nonzero integer Chern number $\mathcal{C}=2$ (tLO$_{2}$) combining the topological non-trivial FF phase with $\mathcal{C}=-1$ (tFF$_{1}$) on the spin--orbit coupled lattice under Zeeman fields. Here, the Rashba SOC splits the single band into two helicity branches; in addition, Cooper pairs $\Delta_{+}$ and $\Delta_{-}$ come from two helicity bands (see Fig.~\ref{fig-FS}a). When there is an in-plane Zeeman field $h_{x}$, the Cooper pairs from different bands have opposite center-of-mass momenta $\pm Q_{y}$ and different amplitudes ($\Delta_{+}$$\neq$$\Delta_{-}$). Moreover, the competition between the Rashba SOC and Zeeman fields leads to the FF phase and the generalized LO phase without nodes\cite{competing}. On this basis, the out-of-plane Zeeman field $h_z$ can drive the FF phase and the LO phase without nodes into the tFF$_{1}$ and tLO$_{2}$ phases for the over-doped and half-filled cases. Under half filling, $\mathcal{C}=2$ becomes a characteristic property for the tSC states of the lattice-type model. In addition to the tFF$_{1}$ states with $\mathcal{C}=-1$ similar to the tFF with $\mathcal{C}=1$ in the continuum Fermi gas\cite{cqu,zha,yecao}, as a new quantum state of matter, the tLO$_{2}$ state, exhibiting the more stable inhomogeneous superfluid/superconducting order, provides a promising candidate for studying the topological properties of matter and searching for MFs. Of note is the fact that the tLO$_{1}$ state with $\mathcal{C}=1$ can emerge in a small-parameter region of the doped phase diagram. To explore their topological properties, we also present the distribution of the lattice field strength for different Chern numbers, the edge states, and their local density of states (LDOS) spectra for inhomogeneous tSC states, which could be tested in future experiments.

\textbf{Model Hamiltonian.} Here we consider a minimal $2$D lattice model with a Rashba SOC and both in-plane and out-of-plane Zeeman fields, which can be described by an effective Hamiltonian:
\begin{eqnarray}
H\!\!&=&\!\!H_{\textrm{K}}+H_{\textrm{R}}+H_{\Delta}, \nonumber \\
H_{\textrm{K}}\!\!&=&\!\!-t \!\sum_{ i,j, \sigma }\! \psi^{\dagger}_{i\sigma} \psi_{j\sigma}\! + \!\sum_{i, \sigma} \mbox{[}h_{\textrm{z}}\sigma_{z}
+h_{\textrm{x}}\sigma_{x}-\mu \mbox{]} \psi^{\dagger}_{i\sigma} \psi_{i\sigma}, \nonumber \\
H_{\textrm{R}}\!\!&=&\!\! \lambda\! \sum_{i}\mbox{(}\psi^{\dagger}_{i}\sigma_{x}\psi_{i+\hat{x}}+\psi^{\dagger}_{i}\sigma_{y}\psi_{i+\hat{y}}+\textrm{h.c.}\mbox{)}, \label{eq:eqn1_rs} \\
H_{\Delta}\!\!&=&\!\! \!\sum_{i}\mbox{(}\Delta_{i} \psi^{\dagger}_{i\uparrow} \psi^{\dagger}_{i\downarrow} +\Delta^{\ast}_{i} \psi_{i\downarrow} \psi_{i\uparrow}\mbox{)}, \nonumber
\end{eqnarray}
where $\psi^{\dagger}_{i\sigma}$ ($\psi_{i\sigma}$) denotes the creation (annihilation) field operator of the fermionic particle with spin $\sigma\equiv \mbox{(}\!\uparrow \mbox{,} \downarrow\!\mbox{)}$ at site $i$, $\lambda$ is the SOC strength, and $h_{x}$ and $h_{z}$ are the in-plane and out-of-plane Zeeman fields, respectively. The effective Hamiltonian can be achieved by solving the following Bogoliubov-de Gennes (BdG) equations:
\begin{small}
\begin{equation}
\vspace{-0.2cm}
\!\sum_{j}\!%
\begin{pmatrix}
H_{ij\uparrow } & \lambda_{1} & 0 & \Delta _{ij} \\
\lambda_{2} & H_{ij\downarrow } & -\Delta _{ij} & 0 \\
0 & -\Delta _{ij}^{\ast } & -H_{ij\uparrow } & \lambda _{3} \\
\Delta _{ij}^{\ast } & 0 & \lambda_{4} & -H_{ij\downarrow }%
\end{pmatrix}%
\begin{pmatrix}
u_{j\uparrow }^{n} \\
u_{j\downarrow }^{n} \\
v_{j\uparrow }^{n} \\
v_{j\downarrow }^{n}%
\end{pmatrix}%
\!=\!\epsilon_{n}%
\begin{pmatrix}
u_{i\uparrow }^{n} \\
u_{i\downarrow }^{n} \\
v_{i\uparrow }^{n} \\
v_{i\downarrow }^{n}%
\end{pmatrix}%
\label{BdG}
\end{equation}%
\end{small}
where $H_{ij\uparrow }=-t_{ij}-(\mu-h_z)\delta _{ij}$,
$H_{ij\downarrow }=-t_{ij}-(\mu+h_z)\delta _{ij}$,
$\lambda_{m}=\lambda \{[(-1)^{m+1} \delta _{i+e_{x},j}+(-1)^{m}\delta _{i-e_{x},j}]+\varepsilon_{m}[i(\delta _{i+e_{y},j}-\delta _{i-e_{y},j})]-\varepsilon_{m}h_x \delta_{i,j}$ with $m=1, 2$, $\varepsilon_{m}=-1$, and $m=3,4$, $\varepsilon_{m}=1$.
The SC pairing $\Delta _{i}=V\langle c_{i\uparrow}c_{i\downarrow}\rangle $ are determined self-consistently for a fixed chemical potential. In the following calculations, the energy is measured in units of the hopping integral $t$ and $\lambda=0.75$. We consider a very low temperature $T=10^{-3}$ and the pairing interaction $V = 2.25$. Here, we assign different initial values of $\Delta _{i}$ using random numbers, and the calculation is repeated until the relative difference in the order parameter between two consecutive iteration steps is less than $10^{-5}$. For multiple solutions, we compare their corresponding free energies to obtain the most energetically favored state.

To distinguish the FF and LO states, we define the amplitude variation of the order parameter $\sigma _{1}=\sqrt{\sum_{i}(|\Delta _{i}|-\overline{|\Delta |})^{2}/N}$ with $\overline{|\Delta |}=\sum_{i}|\Delta _{i}|/N$ and phase variation $\sigma _{2}=\sqrt{\sum_{i}|\Delta _{i}-\overline{\Delta }|^{2}/N}$ with $\overline{\Delta }=\sum_{i}\Delta _{i}/N$. The normal superfluid phase is characterized by $\overline{|\Delta |}\neq 0$, $\sigma _{1}=\sigma _{2}=0$, $Q_{y}=0$. For the LO phase $\overline{|\Delta |}\neq 0$, $\sigma _{1}\neq 0$, $Q_{y}\neq 0$, whereas for the FF phase $\overline{|\Delta |}\neq 0$, $\sigma _{1}=0$, $\sigma _{2}=\overline{|\Delta |}$, $Q_{y}\neq 0$\cite{competing}. To describe the topological properties of the system, we calculated the Chern number in the hole branches $\mathcal{C}=$ $\sum_{l}\mathcal{C}_{l}$. Here, by using the twisted boundary condition\cite{chern} because the system preserves the translational invariance of the system along the $x$ direction, $\mathcal{C}_{l}=\frac{1}{2\pi }\int dk_{x}d\theta_{y}\Omega _{l}$ is the Chern number of the $l$th band. $\Omega _{l}=-2$Im$\left\langle \frac{\partial \Psi _{l}}{\partial k_{x}}|\frac{\partial \Psi_{l}}{\partial \theta_{y}}\right\rangle $ is the Berry curvature \cite{Xiao}, and $\left\vert \Psi _{l}\right\rangle $ is the eigenstate of the $l$th hole band of Eq.~(\ref{BdG}). For the $2$D lattice model, $\mathcal{C}$ can be expressed as the sum of the lattice field strength: $\mathcal{C}=\sum_{n}\frac{\Omega _{n}}{2\pi}$.

\textbf{Pairing mechanism and order parameter}. To distinguish different roles for the SOC and the Zeeman field, we show the Fermi surface for SOC without Zeeman fields (see Fig.~\ref{fig-FS}a). Because of the SOC, the band of the system is split into upper and lower helicity branches, the Fermi surfaces of which are symmetric in the entire Brillouin zone (BZ). Pairing on different Fermi surfaces involves the BCS pairing mechanism between states of opposite momenta $\textbf{k}$ and $-\textbf{k}$. When applying in-plane and out-of-plane Zeeman fields (see Fig.~\ref{fig-FS}b), two helicity Fermi surfaces become asymmetric along the $y$ direction; the upper helicity Fermi surface moves along positive $k_{y}$, and the lower helicity Fermi surface moves along negative $k_{y}$. The asymmetric Fermi surface leads to a finite center-of-mass momentum $\pm Q_{y}$ of the superfluid/superconducting pairings, because the finite momentum Cooper pairs are LO states (with spatially varying order parameter amplitudes). Meanwhile, from the half filling in Fig.~\ref{fig-FS}b, we find that the upper-helicity and lower-helicity Fermi surfaces have the same size, implying that $\Delta_{+}=\Delta_{-}$. As the system deviates from half filling (see Fig.~\ref{fig-FS}c), the two helicity Fermi surfaces have different sizes, implying that $\Delta_{+}$ $\neq$ $\Delta_{-}$. When the system corresponds to the over-doped case (see Fig.~\ref{fig-FS}d), the lower helicity Fermi surface completely disappears, and the topology of the whole Fermi surface is completely changed. Because the contribution of the Cooper pairs comes only from an upper helicity Fermi surface, the system supports an FF state (only $\Delta_{+}$).

\begin{figure}[h!]
\centering
\resizebox{3.4in}{1.7in}{\includegraphics{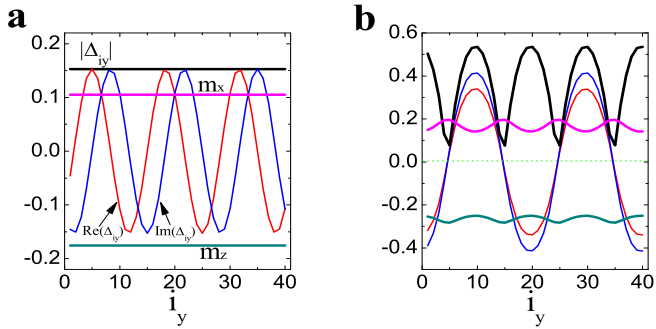}}
\caption{\textbf{Spatial profiles of order parameters for the tFF$_{1}$ and tLO$_{2}$ states}: \textbf{(a)} the FF state order parameter, magnetization $m_{x}$ along the $x$ direction, and $m_{z}$ along the $z$ direction with $\mathcal{C}=-1$, $\mu=-2.95$, $h_{x}=0.6$; \textbf{(b)} order parameter and magnetization for the LO state, $\mathcal{C}=2$, $h_{x}=0.6$, $h_{z}=0.8$.}
\label{fig-order}
\end{figure}

To clarify the difference between the FF and LO states, Fig.~\ref{fig-order} summarizes the detailed spatial profiles of the inhomogeneous tSC order parameters, such as the FF state with spatially uniform magnetization: $m_{x}$ and $m_{z}$ (see Fig.~\ref{fig-order}a), and the LO state with spatially oscillating magnetization, $m_{x}$ and $m_{z}$ (see Fig.~\ref{fig-order}b). For the FF state, the pairing mainly comes from only one of the helicity Fermi surfaces, whereas for the LO state, the pairing comes from both helicity Fermi surfaces. As mentioned previously, compared with the traditional LO state, the real and imaginary parts of the order parameter in the tLO$_{2}$ phase stem from the relative phase arising from the superposition of $\Delta_{+}$ and $\Delta_{-}$, so $\Delta _{i}$ does not have nodes, and the system is gapped completely. Meanwhile, $h_{z}$ drives the system from a non-topological to a topological phase, and we find that the Chern numbers are $-1$ for tFF$_{1}$ and $2$ for tLO$_{2}$.

\begin{figure}[h!]
\centering
\includegraphics[width=3.3in]{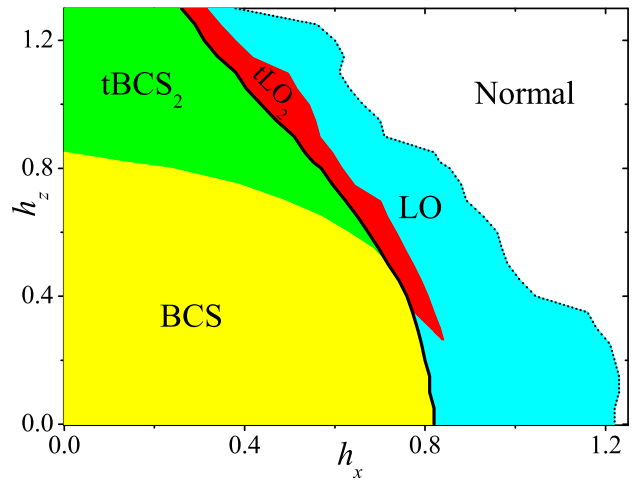}
\caption{\textbf{Phase diagram of the FFLO superfluid for half-filling}. Color identifies the different phases. The superfluid has topological superfluid/superconducting phases with $\mathcal{C}=-1$ (tBCS$_{1}$) and $\mathcal{C}=2$ (tBCS$_{2}$), a LO phase with $\textit{C}=2$ (tLO$_{2}$), a generalized topological trivial LO phase (LO), and a normal phase. Above the dotted line, the order parameter is less than 0.001, corresponding to the normal phase. The solid black line corresponds to the first-order phase transition boundary. Here, we choose $\mu=0.0$.}
\label{fig-phase-half}
\end{figure}

\textbf{Global phase diagrams}. The interplay among the in-plane and out-of-plane Zeeman fields, pairing, and SOC may result in distinct pairing states. To understand these phases better, the global phase diagram is present at half-filling on the $h_{x}-h_{z}$ plane (Fig.~\ref{fig-phase-half}), and there are four different phases, including the conventional BCS state, the topological BCS state with $C=2$, the LO phase, and the tLO$_{2}$ phase. It is well known that the conventional BCS state exists in a region with small $h_{x}$ and $h_{z}$. For $h_{z} < 0.2$, with increasing $h_{x}$, the system is driven from the BCS phase into the generalized LO phase via a first-order transition; finally, $h_{x}$ destroys the SC gap, and the system evolves into a normal metallic state. All these results are consistent with previous non-topological theoretical work. For $0.2 \leq h_{z} \leq0.8$, the LO phase emerges in the large $h_{x}$ region. In this region, the main change is that the BCS phase is initially driven into the tLO$_2$ phase by $h_{x}$ and $h_{z}$, which is a new quantum state of matter. With increasing $h_{x}$, the tLO$_2$ phase transforms into the LO phase. For large $h_{z}\geq 0.8$, the phase structure becomes rich and the topological properties of the system are dominant. Besides the LO phase and the normal phase, there exists a topologically non-trivial tBCS$_{2}$ phase, despite the extremely small $h_{x}$. By turning up the strength of $h_{x}$, the tLO$_{2}$ phase emerges near the region of the tBCS$_{2}$ phase. As $h_{x}$ further increases, the whole system is driven from the tLO$_{2}$ phase into the LO phase. Herein, this new tLO$_{2}$ phase can be simply considered the superposition of two effective \textit{p}-wave pairings (tFF) in two helicity bands.

\begin{figure}[h!]
\centering
\includegraphics[width=3.6in]{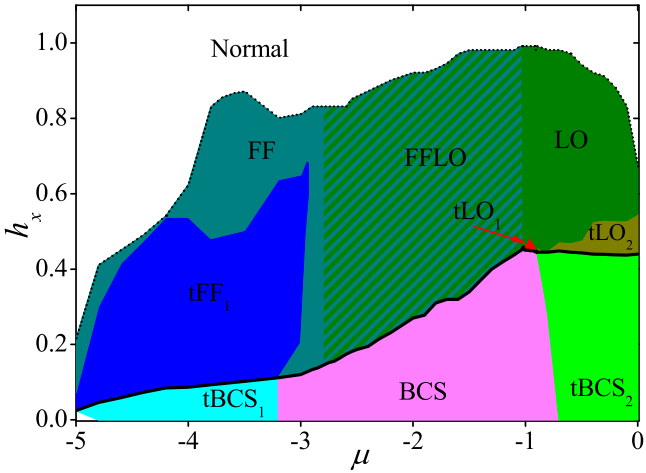}
\caption{\textbf{Phase diagram of a doped FFLO superfluid}. The different phases are labeled with different colors. Compared with half filling, the additional phases are the LO phase with $\mathcal{C}=1$(tLO$_{1}$) and the FF phase with $\mathcal{C}=-1$(tFF$_{1}$). The solid and dotted lines have the same significance as in Fig.~\ref{fig-phase-half}. Here, we choose $h_{z}=1.0$.}
\label{fig-phase}
\end{figure}

\begin{figure}
\centering
\resizebox{3.3in}{3.5in}{\includegraphics{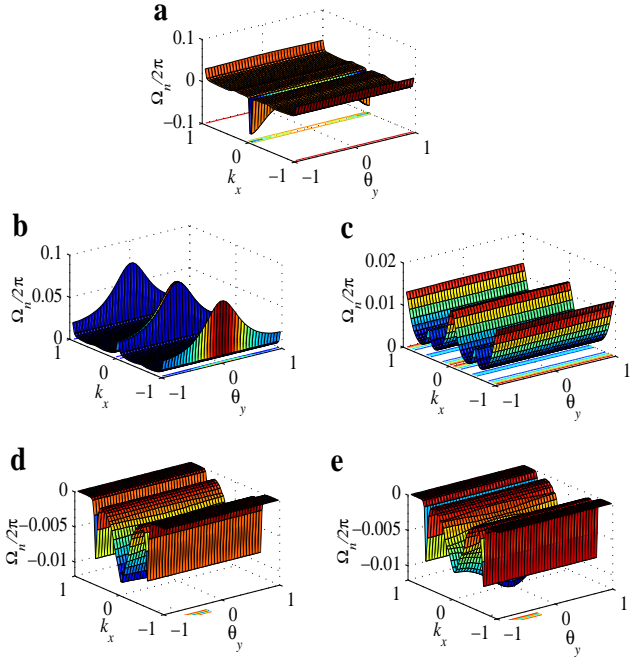}}
\caption{\textbf{Lattice field strength of the Chern numbers for the inhomogeneous tSC states}. Berry curvature of the tLO$_{1}$ state with $\mathcal{C}=1$, $\mu = -0.93$, $h_{x}=0.47$ for \textbf{(a)}; \textbf{(b)} the tBCS$_{2}$ state with $\mathcal{C}=2$, $h_{z}=0.8$, $h_{x}=0.3$; \textbf{(c)} the tLO$_{2}$ state with $\mathcal{C}=2$, $h_{z}=0.8$, $h_{x}=0.6$; \textbf{(d)} the tBCS$_{1}$ state with $\mathcal{C}=-1$, $\mu=-3.2$, $h_{x}=0.1$; \textbf{(e)} the tFF$_{1}$ state with $\mathcal{C}=-1$, $\mu = -2.95$, $h_{x} = 0.6$.}
\label{fig-berry}
\end{figure}

The $\mu$--$h_{x}$-plane phase diagram is presented for a system far from half filling through chemical doping; see Fig.~\ref{fig-phase}. In the region close to half filling, we obtain the tBCS$_{2}$ phase, the tLO$_{2}$ phase, and the LO phase, which is consistent with half filling. With decreasing $\mu$, the topological properties of the tBCS$_{2}$ phase disappear, and the tBCS$_{2}$ phase evolves into a conventional BCS phase for small $h_{x}$; for moderate $h_{x}$, with decreasing chemical potential, the $\mathcal{C}$ of the tLO$_{2}$ phase changes from $2$ to $1$, and the system accordingly enters into the tLO$_{1}$ phase. From Fig.~\ref{fig-berry}a, the topological property of the tLO$_{1}$ phase principally comes from the contribution of the upper helicity Fermi surface including points of $k_{x}=\pi$. For large $h_{x}$, when the chemical potential is in the interval $-2.8\leq \mu \leq1$, distinguishing the FF and LO phases is difficult because the two are very close, so they are collectively called the FFLO phase. When the system is far from half filling $\mu\leq-2.8$, there exists only one helicity Fermi surface. In this circumstance, the tBCS$_{1}$ phase emerges for small $h_{x}$. With increasing $h_{x}$, the tBCS$_{1}$ phase evolves to the tFF$_{1}$ phase within the FF phase. Here, $\mathcal{C}$ comes from the contribution of the upper helicity Fermi surface for both tLO$_{1}$ and tFF$_{1}$ phases. Although the upper helicity Fermi surface contributes to the Chern numbers of the tLO$_{1}$ and tFF$_{1}$ phases, the topology of their Fermi surfaces is completely different, corresponding to different levels of chemical doping, as seen in Fig.~\ref{fig-FS}c and d.

\begin{figure}[h!]
\centering
\resizebox{3.5in}{3.2in}{\includegraphics{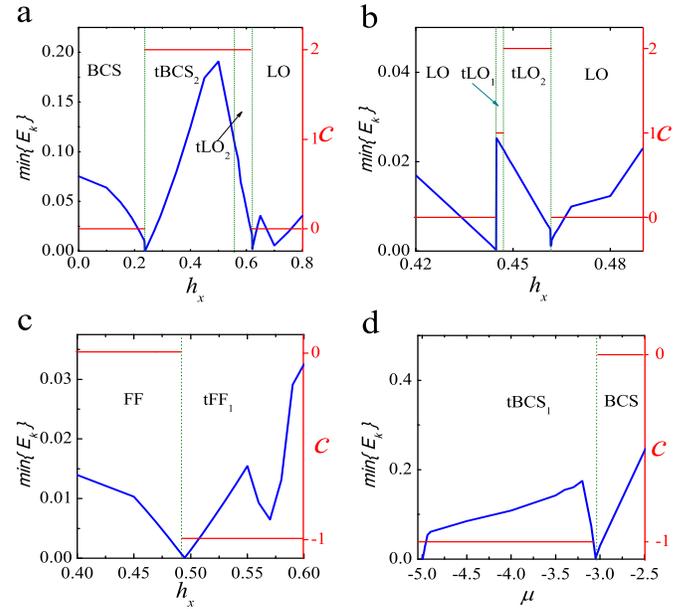}}
\caption{\textbf{Evolution of minimal excitation gap of inhomogeneous tSC states}.
Evolution of the minimal excitation gap across the topological phase boundary with $h_{x}$ in \textbf{(a)}, \textbf{(b)}, and \textbf{(c)}; in \textbf{(d)}, the topological phase transition emerges with increasing chemical potential. The red line presents the Chern number. When increasing $h_{x}$ and $\mu$, at the topological phase boundary, the excitation closes and reopens. In \textbf{(a)}, $\mu=0$, $h_{z}=0.8$; \textbf{(b)}, $\mu=-0.9$, $h_{z}=1$; \textbf{(c)}, $\mu=-2.95$, $h_{z}=1$; \textbf{(d)}, $h_{z}=1$, $h_{x}=0$.}
\label{fig-close}
\end{figure}

\textbf{Evolution of minimal excitation gap and Chern number}. The band inversion is an important feature of the emergence of a topological quantum transition, and this inversion corresponds to closing and reopening the minimal excitation gap $E_{g}$. To understand the topological transition better, we present the evolution of a minimal excitation gap $E_{g}$ across the topological phase boundary with $h_{z}$ and $\mu$ in Fig.~\ref{fig-close}. For a fixed out-of-plane Zeeman field $h_{z}=0.8$ such as in Fig.~\ref{fig-close}a, with increasing $h_{x}$, at a fixed point of $h_{x}$, $E_{g}$ first closes and reopens, marking the transition from the BCS phase into the tBCS$_{2}$ phase, and the Chern number changes from $0$ to $2$, accordingly. With a further increase in $h_{x}$, the tBCS$_{2}$ phase turns into the tLO$_{2}$ phase. With a continued increase in $h_{x}$, at the second fixed point of $h_{x}$, $E_{g}$ closes and reopens again, and the Chern number changes from $2$ to $0$, accordingly, marking the transition from the tLO$_{2}$ phase into the LO phase. For $h_{z}=1$ and $\mu=-0.9$(Fig.~\ref{fig-close}b), because $h_{z}$ is stronger, the system first presents the LO phase, and then transforms into the tLO$_{1}$ phase($\mathcal{C}=1$) with increasing $h_{x}$. At the transition point, we detect a closing and reopening of $E_{g}$. At the same time, the Chern number becomes 2. Nevertheless, this tLO$_{1}$ phase extends over a small domain and with increasing $h_{x}$, it quickly enters into the tLO$_{2}$ phase. With stronger $h_{x}$, the order parameter begins to become progressively smaller, and increasing numbers of nodes begin to emerge. These nodes destroy the topological properties of the system and drive the system from the tLO$_{2}$ into the LO phase again. For a sufficiently low band filling (Fig.~\ref{fig-close}c), we can obtain further access to the FF phase, when $h_{x}$ is increased; the FF phase transforms into the tFF$_{1}$ with $\mathcal{C}=-1$. Overall, for a sufficiently strong $h_{x}$, the system finally enters the normal phase. In addition to $h_{x}$, by tuning down band filling (see Fig.~\ref{fig-close}d), a topological phase transition also emerges between the BCS and tBCS$_{1}$ phases. We summarize these results by concluding that changes in $h_{x}$ and adjustments of $\mu$ yield a variety of different topological phases for the system.

\begin{figure}
\centering
\resizebox{3.4in}{4.4in}{\includegraphics{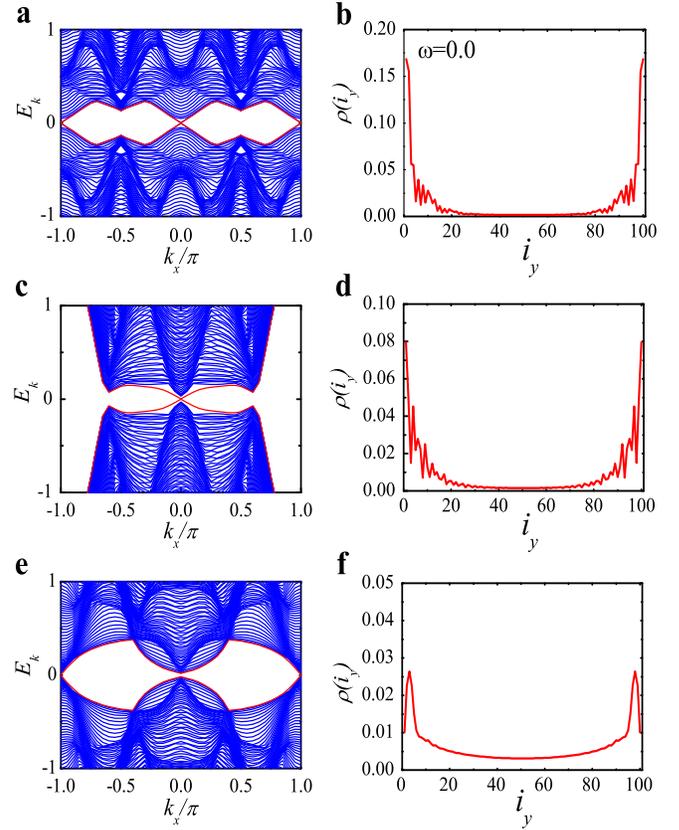}}
\caption{\textbf{Edge states and local density of states for tLO$_{2}$, tFF$_{1}$, and tLO$_{1}$ phases in a $2$D strip}: \textbf{(a)} the edge state of the tLO$_{2}$ with $C=2$, $h_{z}=0.8$ $h_{x}=0.6$; \textbf{(b)} the local density of states for the tLO$_{2}$ state at zero energy, with the same parameters as those in \textbf{(a)};
\textbf{(c)} the edge state of the tFF$_{1}$ with $C=-1$, $\mu=-3$, $h_{x}=0.4$; \textbf{(d)} the local density of states for the tFF$_{1}$ state, with the parameters as those in \textbf{(c)}; \textbf{(e)} the edge state of the tLO$_{1}$ with $C=1$, $\mu=-0.93$, $h_{x}=0.47$,$h_{z}=1.0$; \textbf{(f)} the local density of states for the tLO$_{1}$ state, with the parameters as those in \textbf{(e)}.}
\label{fig-edge}
\end{figure}

To clearly show the origin of the Chern number in these tSC states, we calculated the lattice field strength of the Chern number (see Fig.~\ref{fig-berry}). For the tBCS$_{1}$ state in Fig.~\ref{fig-berry}a, we see that both peak streaks in field strength point upwards and combine constructively with the positive background, to contribute to a nonzero Chern number $1$. In contrast, a downward-pointing peak streak arises around $k_{x}=0$ and does not contribute to a nonzero Chern number because it opposes the positive background. For the tBCS$_{2}$ and tLO$_{2}$ states of Fig.~\ref{fig-berry}b and c, three positive-peak streaks in the lattice field strength point upwards and combine constructively. Hence, the non-zero Chern number is determined by both the non-trivial points around $k_{x}=0$ and $k_{x}=\pi$, which belong to the high and low helicity Fermi surfaces, respectively. Compared with the tBCS$_{2}$ and tLO$_{2}$ states, the main contribution of $\mathcal{C}$ for the tBCS$_{1}$ and tFF$_{1}$ states comes from three streaks of negative peaks near $k_{x}=0$, which means $\mathcal{C}=-1$ is determined by the non-trivial points near $k_{x}=0$, belonging to only one helicity Fermi surface. Clearly, the value of $\mathcal{C}$ is closely related to the topological structure of the two helicity Fermi surfaces. We know the two topological points $k_{x}=0$ and $k_{x}=\pi$ will be highly critical in the lattice system, as reflected by the edge states in Fig.~\ref{fig-edge}. In these edge states, gapless edge modes arise at either $k_{x}=0$, $k_{x}=\pi$, or two simultaneous points in pairs. Hence the value of $\mathcal{C}$ depends on comprehensive effects of the special structure of the two helicity Fermi surfaces at the half-filling and doping level.

\textbf{Edge state and the LDOS}. For the tFF($\mathcal{C}=1$) state, the preceding theoretical work shows support for exotic chiral edge modes\cite{cqu,zha,yecao}. In this work, we looked into the question of whether the tLO$_{1,2}$ and tFF$_{1}$ states support exotic chiral edge modes. To understand the topological property of the tLO$_{1,2}$ and tFF$_{1}$ states more clearly, we consider a $2$D strip, the lattice size of which has width $x=40$ and length $y=100$. Because the state obeys translation invariance along the $x-$direction, we adopt periodic and open boundary conditions along the $x-$ and $y-$directions, respectively, with $k_{x}$ as a good quantum number. The energy spectrum and the LDOS are plotted in Fig.~\ref{fig-edge}. There exists a distinct energy gap because of the SOC, and gapless edge states appear at $k=0$ and $k=\pi$; see Fig.~\ref{fig-edge}a. For the tFF$_{1}$ state, a gapless edge state appears only at $k=0$ (Fig.~\ref{fig-edge}c), and only at $k=\pi$ for the tLO$_{1}$ state (Fig.~\ref{fig-edge}e). The presence of edge states indicates topological properties for both the tLO$_{1,2}$ and tFF$_{1}$ states in momentum space. In real space, we calculated the LDOS spectra at $\omega=0$ along the lattice $i_{y}$ in Fig.~\ref{fig-edge}b, d, and f. At the edge of the lattice system, we find two zero-energy modes, which implies that the tLO$_{2}$ and tFF$_{1}$ states support two local MFs at two edges of the strip. For the tLO$_{1}$ state, the gap protecting the edge state is quite narrow, and the scattering strength from the edge becomes relatively strong, which leads to the formation of the Mott insulating gap at the edge of the strip. The Cooper pairs in bulk bounce off from the Mott insulating gap and form a new edge state in the new location. This results in migration of the edge state towards the center of the bulk. Thus, we find that the edge state of the tLO$_{1}$ is not exactly at the two edges of the strip, but near the center of the bulk (Fig.~\ref{fig-edge}f).

Our proposed tLO$_{2}$ and tFF$_{1}$ phases may also be realized in a spin-orbit coupled lattice system of semiconductor/superconductor heterostructures or cold atom gas\cite{JSau,Roman}. In particular, for cold-atom systems, some promising theoretical proposals to generate Rashba-type SOCs have already been proposed\cite{liu}, and the effective Zeeman field and SOC have been realized experimentally by two-photon detuning and modulated Raman fields \cite{soc1,soc2,soc3}. All this theoretical and experimental progress has laid a solid foundation to achieve inhomogeneous tSC states. Our numerical results are rather different from the results using a continuum model, and the lattice structure plays an important role in obtaining and stabilizing these exotic quantum states, especially the LO and tLO$_{2}$ states, which appear more stable. As a first attempt, we predict new quantum states---the tFF$_{1}$ state and the tLO$_{1,2}$ state---under the present lattice model, and their emergence also uncovers a novel mechanism for inhomogeneous tSC states in two spin-mixed asymmetric helicity bands of the SOC system. Compared with the topological trivial FFLO states, the topological FFLO states might be realized in experiments more easily through topological protection.

In summary, we propose that tFF$_{1}$ and tLO$_{1,2}$ states with finite momentum pairings can be realized using the SOC lattice under Zeeman fields with $s$-wave pairing. A global phase diagram has been presented, which includes different topological phases in the given parameter region. We also explored the detailed structures of the inhomogeneous order parameters, the distribution of the lattice field strength for different Chern numbers, the edge states of the tSC phases, and LDOSs. The tLO$_{2}$ state is identified for the first time in the SOC lattice under Zeeman fields and combines with the tFF$_{1}$ state to broaden and deepen our understanding of topological superfluids.

\textbf{Acknowledgements.} We thank Y.S. Wu, R.B. Tao, T.K. Lee, and A. Varlamov for fruitful discussions. This work was supported by the State Key Programs of China (Grant No. 2017YFA0304204£¬and 2016YFA0300504) and the National Natural Science Foundation of China (Grant Nos. 11625416 and 11474064).

\end{document}